\documentclass[apjl]{emulateapj}
\usepackage{natbib}

\newenvironment{myitemize}
{
    \begin{list}{$\bullet$}{}
        \setlength{\itemsep}{0pt} 
}
{
    \end{list} 
}

\newcommand{\myemail}{pedro@ifa.hawaii.edu}
\newcommand{\tna} {\,\tablenotemark{a}}
\newcommand{\tnb} {\,\tablenotemark{b}}
\newcommand{\tnc} {\,\tablenotemark{c}}
\newcommand{\tnd} {\,\tablenotemark{d}}
\newcommand{\tne} {\,\tablenotemark{e}}
\newcommand{\tnf} {\,\tablenotemark{f}}
\newcommand{\tng} {\,\tablenotemark{g}}

\newcommand{\kgm} {$\,$kg$\,$m$^{-3}$}

\shorttitle{Photometry of 2003 EL61}
\shortauthors{Lacerda, Jewitt, Peixinho}
\slugcomment{Accepted for publication in The Astronomical Journal on 2008/01/25}

\begin{document}

\def\FigSpectrum{
   \begin{figure}
   \centering
      \includegraphics[width=1.0\columnwidth]{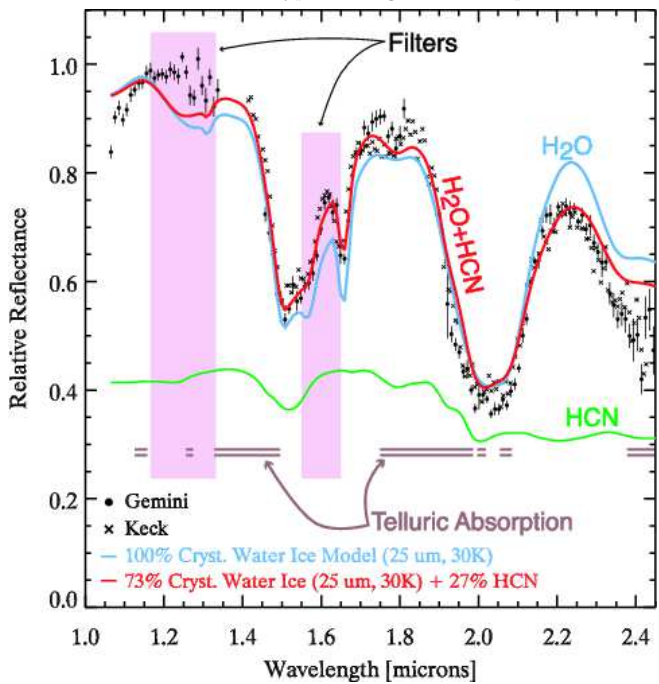}

   \caption[f1.eps] {Near IR Spectrum of 2003 EL61, adapted from Trujillo et
al. (2007). A pure crystalline water-ice model fit and a mix of water ice and
HCN ice are overplotted. The locations and approximate widths of the
1.25$\,\mu$m and 1.6$\,\mu$m filters used to monitor the 1.5$\,\mu$m water-ice
band depth, as well as the wavelength regions where the Earth's atmosphere is
opaque, are also shown.} 

   \label{Fig.Spectrum}
   \end{figure}
}

\def\FigBRJLightcurve{
   \begin{figure*}
   \centering
      \includegraphics[width=0.9\textwidth]{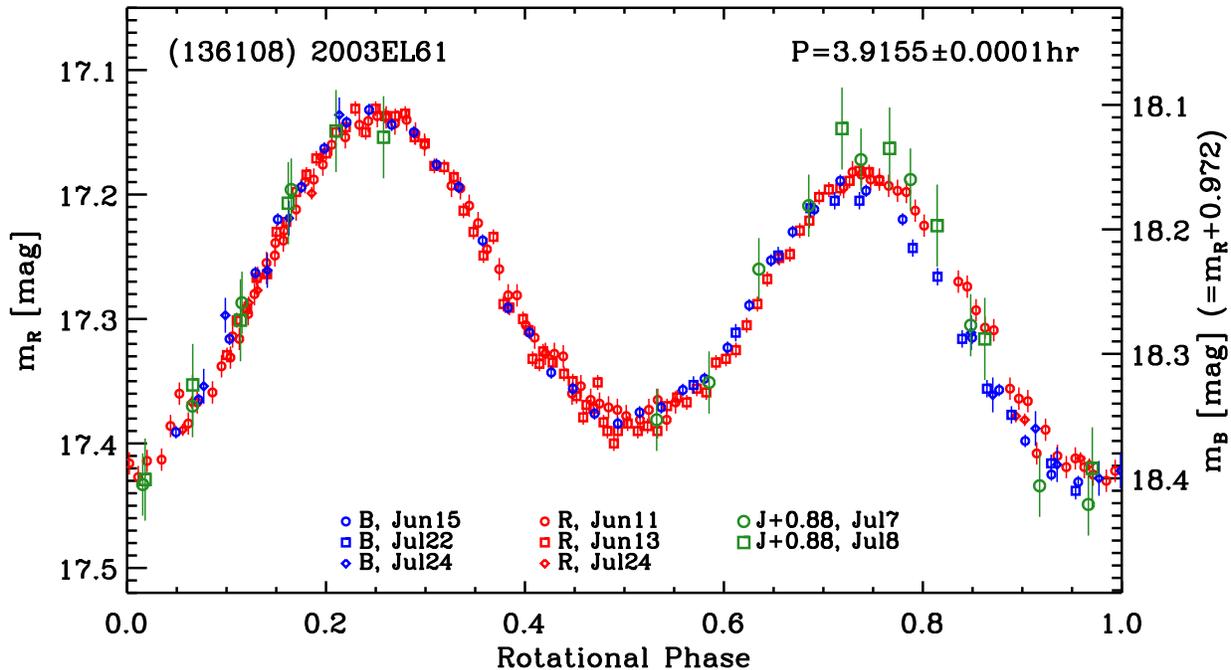}

   \caption[f2.eps] {Lightcurves of EL61 through the $B$-, $R$-, and $J$-band.
$B$ and $J$ data were scaled to the $R$ data by subtracting the median colors
$V-R=0.972$ and $R-J=0.88$. The error bars are $\sim0.01\,$mag.\ in the $B$ and
$R$ bands, and $\sim0.03\,$mag.\ in the $J$ band.} 

   \label{Fig.BRJLC}
   \end{figure*}
}

\def\FigBRColorCurve{
   \begin{figure}
   \centering
      \includegraphics[width=1.0\columnwidth]{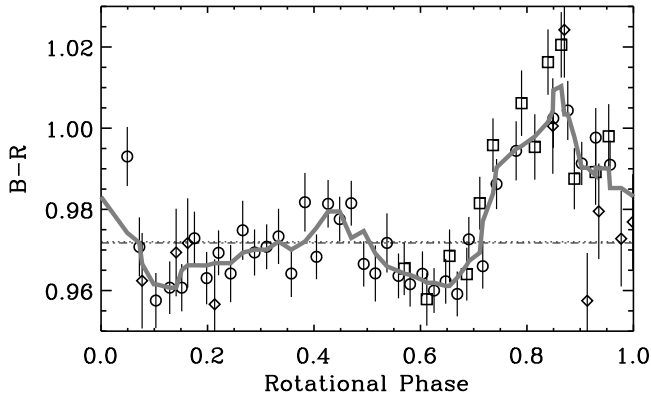}

   \caption[f3.eps] {Circles, squares, and diamonds mark the difference between
the $B$ data and $R$ data interpolated at the $B$ rotational phases. Different
symbols indicate different nights. A running median (width=6) is overplotted as
a thick gray line. A horizontal dotted and dot-dashed lines respectively mark
the mean and median color. The reddening in the region from 0.70 to 1.05 in
rotational phase is clear.  } 

   \label{Fig.BRColorCurve}
   \end{figure}
}

\def\FigSpotCurves{
   \begin{figure}
   \centering
      \includegraphics[width=1.0\columnwidth]{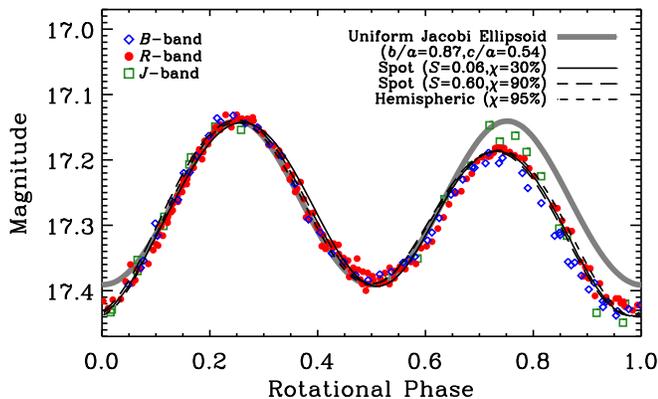}

   \caption[f4.eps] {$B$, $R$ and $J$ lightcurves of EL61 with four models
overplotted.  The thick grey line corresponds to a Jacobi equilibrium ellipsoid
model (axis ratios $b/a=0.87$ and $c/a=0.54$), assumed to have uniform surface
optical properties.  The three thin black lines correspond to models with
non-uniform surfaces. ``Spot'' models have darker circular regions located on
the equator of the Jacobi ellipsoid, leading a semi-major axis by 45\degr. The
numbers in parenthesis indicate the area ($S$, relative to the maximum
cross-section of the ellipsoid, $\pi ac$) and albedo ($\chi$, relative to the
surrounding regions) of the spot.  In ``Hemispheric,'' the darker region covers
a whole hemisphere of EL61. } 

   \label{Fig.JacSpotCurves}
   \end{figure}
}

\def\FigWaterBandDepth{
   \begin{figure}
   \centering
      \includegraphics[width=1.0\columnwidth]{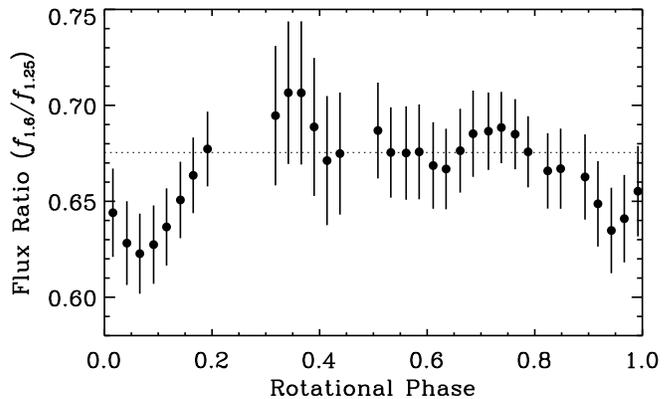}

   \caption[f5.eps] {The ratio of the flux density at 1.6$\,\mu$m to the
continuum flux density at 1.25$\,\mu$m measured on UT 2007 July 07. A thin
horizontal dotted line marks the median of the data points. See text for
details. } 

   \label{Fig.WaterBandDepth}
   \end{figure}
}

\def\FigSpotModels{
   \begin{figure*}
   \centering
      \includegraphics[width=0.9\textwidth]{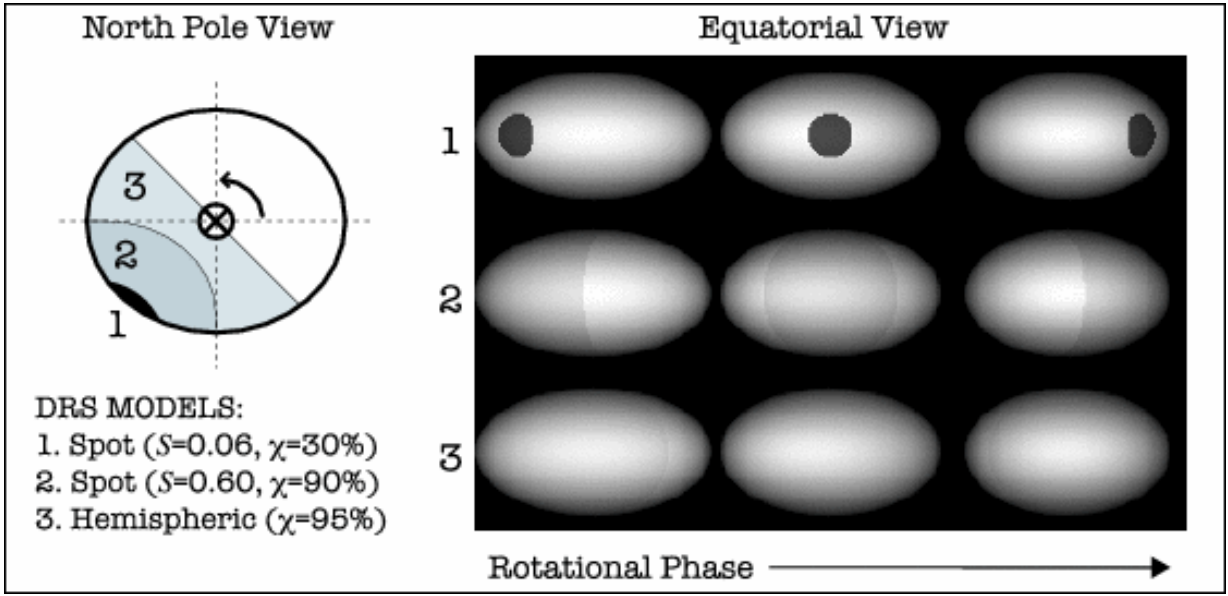}

   \caption[f6.eps] {Sample spot models used to fit the lightcurve data
of EL61 in Fig.\ref{Fig.JacSpotCurves}. North pole and three equatorial views
of the ellipsoid (from left to right: flank-on, spot-on, and tip-on, or
rotational phases $\sim$0.750, $\sim$0.875, and $\sim$1.000 in
Fig.~\ref{Fig.JacSpotCurves}) are shown.  The spot in each model is
characterized by a surface area $S$ (expressed as a fraction of the maximum
equatorial cross-sectional area of EL61) and an albedo, $\chi$, normalized to
the albedo of the surface outside the spot.  The spots are assumed to be
located on the equator of EL61 and leading a semi-major axis by 45\degr.
``Hemispheric'' is a model in which a whole hemisphere of EL61 has a darker
albedo.  The albedo ratio $\chi=95$\% in the hemispheric model (3) is almost
imperceptible.  See text for details.} 

   \label{Fig.SpotModels}
   \end{figure*}
}

\def\FigAlbedoAndColorVsArea{
   \begin{figure}
   \centering
      \includegraphics[width=1.0\columnwidth]{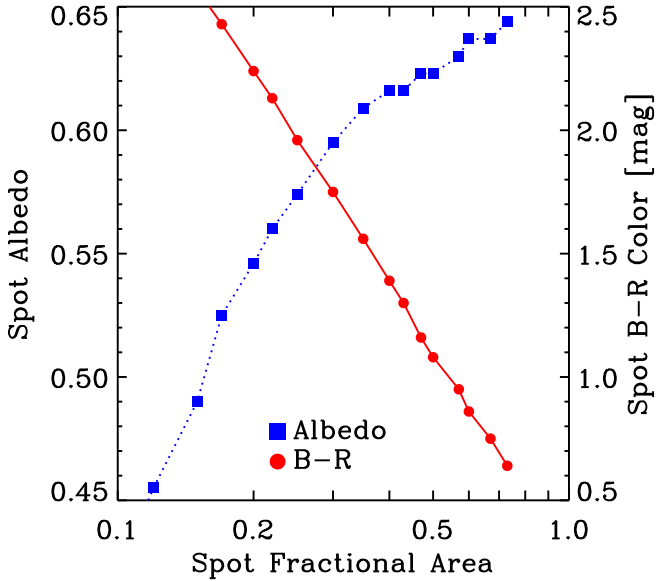}

   \caption[f7.eps] {Range of models consistent with the lightcurve data.
Plotted on the left vertical axis is the assumed albedo of the spot material
(the average geometric albedo of EL61 is $p=0.70$) while on the right vertical
axis we plot the assumed $B-R$ color index of the spot. The horizontal axis
shows the area of the spot (as a fraction of the maximum projected
cross-section of the best-fit equilibrium figure, $\pi ac$.) } 

   \label{Fig.AlbedoAndColorVsArea}
   \end{figure}
}

\def\FigColorVsAlbedo{
   \begin{figure}
   \centering
      \includegraphics[width=1.0\columnwidth]{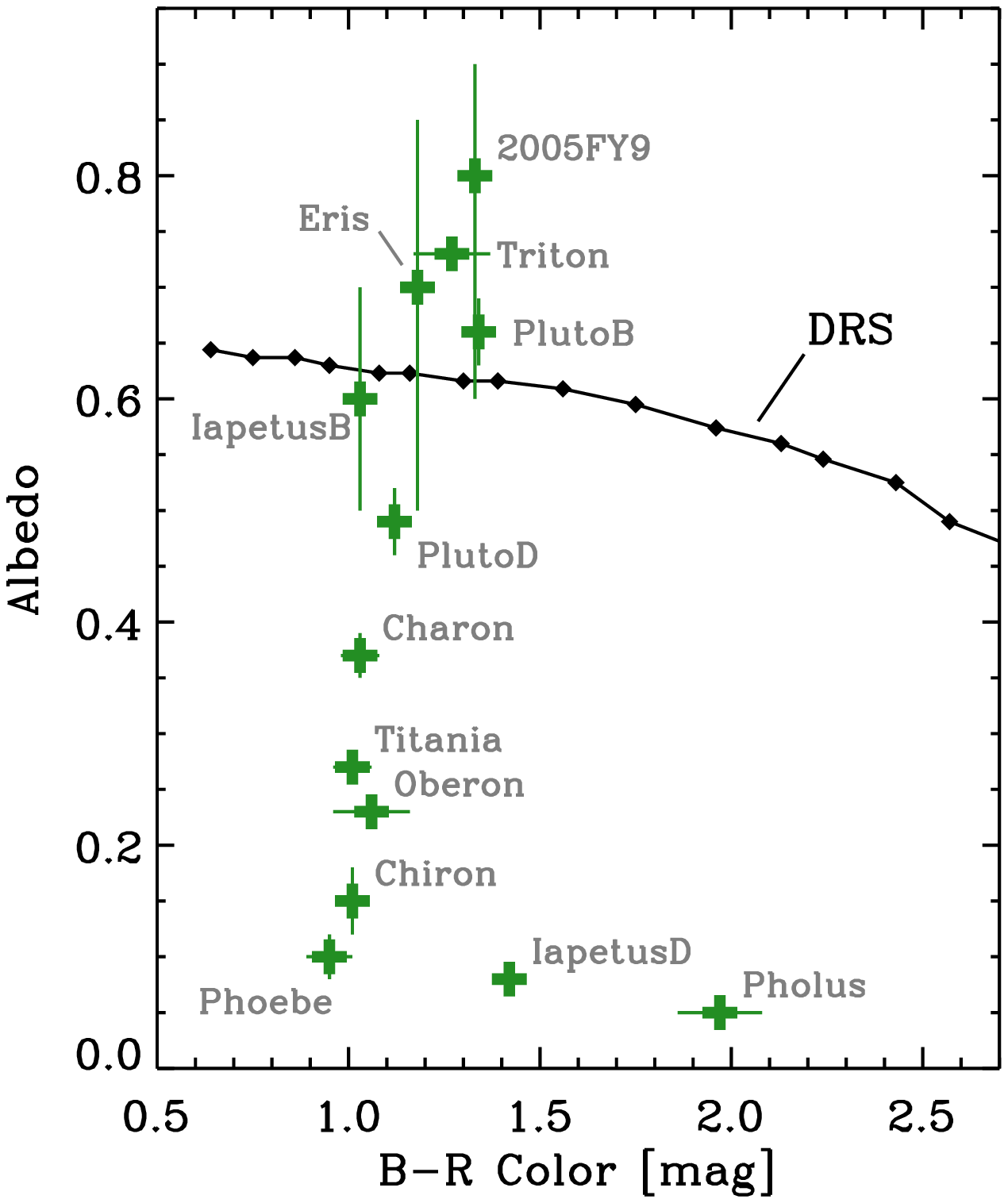}

   \caption[f8.eps] {Combined constraints from Fig.
\ref{Fig.AlbedoAndColorVsArea} on the albedo and $B-R$ color of the DRS.
Overplotted are the albedo, color pairs of identified outer Solar system
surface types. For Iapetus and Pluto, objects with large albedo contrasts, the
labels B and D correspond to the bright and dark areas, respectively.}

   \label{Fig.ColorVsAlbedo}
   \end{figure}
}

\title{High Precision Photometry of Extreme KBO 2003$\,$EL$_{61}$}

\author{Pedro Lacerda$^1$}

\author{David Jewitt$^1$}


\author{Nuno Peixinho$^{1,2}$}

\affil{1 Institute for Astronomy, University of Hawaii, 2680 Woodlawn
Drive, Honolulu, HI 96822}
\affil{2 Grupo de Astrof{\'\i}sica, Universidade de Coimbra, Portugal}

\email{\myemail}

\begin{abstract}

We present high precision, time-resolved visible and near infrared photometry
of the large (diameter $\sim$ 2500 km) Kuiper belt object (136108)
2003$\,$EL$_{61}$.  The new data confirm rapid rotation at period $P$ =
3.9155$\pm$0.0001$\,$hr with a peak-to-peak photometric range $\Delta m_R$ =
0.29$\pm$0.02$\,$mag.\ and further show subtle but reproducible color variations
with rotation.  Rotational deformation of 2003$\,$EL$_{61}$ alone would give
rise to a symmetric lightcurve free of color variations. The observed
photometric deviations from the best-fit equilibrium model show the existence
of a large surface region with an albedo and color different from the mean
surface of 2003$\,$EL$_{61}$.  We explore constraints on the nature of this
anomalous region set by the existing data.

\end{abstract}


\section{Introduction}

The known Kuiper belt objects (KBOs) extend in size from bodies so small as to
be at the limits of sensitivity of the largest telescopes up to Pluto-sized
bodies large enough to have body shapes controlled by self-gravity.  The large
objects are particularly amenable to physical study and a number of intriguing
results have already been secured.  The case of Pluto is well known: the main
object rotates slowly (period $\sim$6 days) but the massive satellite Charon
carries enough angular momentum that the system as a whole is near the critical
threshold for breakup \citep{1989ApJ...344L..41M}.
Surface maps derived from mutual occultation events show a spatially variegated
surface, with a wide range of local albedos \citep[from 0.1 to
0.9:][]{1992Icar...97..211B,1999AJ....117.1063Y}
that may be related to surface deposition of frosts from Pluto's thin
atmosphere \citep{1989GeoRL..16.1213T}.  
Other KBOs have less well known physical properties but new data are beginning
to reveal a startling range of surface types
\citep{2004Natur.432..731J,2007AJ....133..526T,2007ApJ...655.1172T}
and rotational \citep{2006AJ....131.2314L,2007AJ....134..787S} properties.
Notable examples of the latter include $\sim$900 km diameter (20000) Varuna,
whose 6$\,$hr period and 0.4$\,$mag.\ photometric range are best explained as
products of a rotationally deformed body shape and a bulk density of 1000\kgm\
\citep{2002AJ....123.2110J,2004PASJ...56.1099T,2007AJ....133.1393L}.
The large amplitude (1.14$\pm$0.04 mag.), long rotation period
(13.7744$\pm$0.0004$\,$hr) and eclipsing binary-like lightcurve of
$\sim$240$\,$km diameter 2001$\,$QG$_{298}$ suggest an even more extreme
interpretation as a contact or near-contact binary
\citep{2004AJ....127.3023S,2007AJ....133.1393L}.   

One of the most remarkable of the large KBOs yet to be identified is (136108)
2003$\,$EL$_{61}$ (hereafter ``EL61''), whose rapid rotation
(period$\sim$3.9154$\pm$0.0002$\,$hr), and lightcurve range ($\Delta m_R \sim
0.4\,$mag) and near-symmetric morphology together suggest a rotationally
deformed body of density $\sim$2500\kgm\
\citep{2006ApJ...639.1238R,2007AJ....133.1393L}.
EL61 is further interesting in its own right, as an extreme example of a large
KBO with a rapid spin and also as the possible parent of a reported dynamical
cluster of KBOs, perhaps produced by an ancient, shattering collision (Brown et
al.  2007, Ragozzine and Brown 2007). Some members of this dynamical cluster
share spectral features with EL61. Nearly all Pluto-sized KBOs have methane
rich surfaces. EL61 is unusual in that it is covered in almost pure crystalline
H$_2$O ice (see Fig.~\ref{Fig.Spectrum}; Trujillo et al. 2007).  In this paper,
we present new high-precision, time-resolved photometry taken to further
explore the nature of EL61.   

\begin{deluxetable*}{cccccccc}
  \tablecaption{Journal of Observations. \label{Table.Journal}}
   \tablewidth{0pt}
   \tablehead{
   \colhead{UT Date} & \colhead{$R$\tna} & \colhead{$\Delta$\tnb} & \colhead{$\alpha$\tnc} & \colhead{Tel.\tnd} & \colhead{Filt.\tne} & \colhead{Seeing\tnf} & \colhead{Exp. Time\tng} \\ 
   \colhead{} & \colhead{$[AU]$} & \colhead{$[AU]$} & \colhead{[\degr]} & \colhead{} & \colhead{} & \colhead{[\arcsec]} & \colhead{[s]} 
   }
   \startdata
2007 Jun 11 & 51.1570 & 50.8037 & 1.07 &UH2.2m & $R$      & 0.9 & 80 \\
2007 Jun 13 & 51.1567 & 50.8296 & 1.08 &UH2.2m & $R$      & 1.0 & 80 \\
2007 Jun 15 & 51.1565 & 50.8576 & 1.09 &UH2.2m & $B$      & 0.9 & 260 \\
2007 Jul 07 & 51.1536 & 51.1800 & 1.14 &UKIRT  & $J$      & 1.0 & 60 \\
2007 Jul 08 & 51.1535 & 51.1949 & 1.14 &UKIRT  & $J$      & 1.2 & 60 \\
2007 Jul 22 & 51.1517 & 51.4001 & 1.10 &UH2.2m & $B$      & 1.5 & 300 \\
2007 Jul 24 & 51.1514 & 51.4286 & 1.09 &UH2.2m & $B$, $R$ & 1.5 & 260, 80 \\
   \enddata
  \tablenotetext{a}{Heliocentric distance in AU;}
  \tablenotetext{b}{Geocentric distance in AU;}
  \tablenotetext{c}{Phase angle in degrees;}
  \tablenotetext{d}{Telescope used;}
  \tablenotetext{e}{Filters used;}
  \tablenotetext{f}{Typical seeing in arcseconds;}
  \tablenotetext{g}{Typical integration time per frame in seconds.}
\end{deluxetable*}

\section{Observations}

Optical observations were taken using the 2.2-m diameter University of Hawaii
telescope atop Mauna Kea, Hawaii.  We used a Tektronix 2048$\times$2048 pixel
charge-coupled device (CCD) mounted at the f/10 Cassegrain focus, giving pixels
each 0.219 arcsecond square.  Observations were obtained through broadband
$BVRI$ filters approximating the Kron-Cousins photometric system.  The data
were instrumentally calibrated using bias frames and flat-field images obtained
from dithered, median-combined images of the twilight sky.  Photometric
calibration was obtained from observations of standards PG1323-085C, 107~457,
Markarian A1, and PG1633-099A from the list by Landolt (1992).

\FigSpectrum

Near infrared observations were taken using the 3.8-m diameter United Kingdom
Infrared Telescope (UKIRT), also located on Mauna Kea.  We used the UIST
imaging camera, which houses a 1024$\times$1024 pixel array having image scale
of 0.12 arcsec per pixel.  Our principal aim was to use the near-infrared
wavelengths to search for rotational variability of water ice on the surface of
EL61.  For this purpose, we elected to use two filters, one to measure the 1.5
$\mu$m band of water ice and the other to sample the reflected continuum.  Use
of two filters, as opposed to a near infrared spectrometer, allowed us to
maintain rapid sampling (important because of the short rotation period of
EL61) and high signal-to-noise ratios.   Given the available UKIRT filter set,
we employed the ``CH4$_s$'' filter (center 1.60$\,\mu$m, full width at half
maximum FWHM = 0.11$\,\mu$m) to measure the water band (see
Fig.~\ref{Fig.Spectrum}).  The Mauna Kea $J-$band filter (center 1.25$\,\mu$m,
FWHM = 0.16$\,\mu$m) provided a suitable measure of the continuum.  In the
remainder of the text we refer to these filters as ``1.6$\,\mu$m'' and
``1.25$\,\mu$m''.  Photometric calibration of the UKIRT data was obtained from
observations of standard stars S791-C and S813-D from Persson et al.  (1998).
The flux through each filter was measured relative to a field star and a second
star was used to verify the regularity of the first. Since simultaneous
measurements through the 1.25$\,\mu$m and 1.6$\,\mu$m filters were not
possible, we cross-interpolated their fluxes to the same times and measured the
ratio of the interpolated values. A summarized journal of observations can be
found in Table~\ref{Table.Journal}, and the final calibrated broadband
photometric measurements are shown in Tables~\ref{Table.BData} through
\ref{Table.JData}.  Table~\ref{Table.wData} shows the ratio of flux density at
1.6$\,\mu$m to the continuum flux density at 1.25$\,\mu$m, as a function of
time.

\section{Results and Discussion}


Photometric measurements were obtained first relative to field stars, to
provide protection from transient changes in the transparency of the Earth's
atmosphere and variations in the seeing (which can impact the accuracy of
photometry obtained through discrete apertures).  The resulting measurements
were calibrated against standard stars using large aperture photometry.
Internal accuracy of the lightcurve data in the $B$ and $R$ filters is good to
about $\pm$0.01$\,$mag.\ while, in the infrared, scatter in the photometry
shows that the accuracy is at the $\pm$0.03$\,$mag.\ level.  We did not apply a
phase angle correction to the data.  The phase angle changed from 1.07 to 1.10
degrees in our $B$ and $R$ observations.  In this 0.03 degree phase angle
range, with a phase coefficient of $\sim$0.1 mag/deg
\citep{2007AJ....133...26R},
the effect of phase is only 0.003$\,$mag.\ and therefore unimportant compared
to the 0.01$\,$mag.\ photometric accuracy.  Furthermore, the color dependence
of the phase coefficient is small for EL61 according to these authors, and the
expected change in the color resulting from phase angle is only about 0.001
mag, which is again negligible.

The best-fit lightcurve period was determined from the $R$-band data using
phase-dispersion minimization (PDM; Stellingwerf 1978) as $P$ =
3.9155$\pm$0.0001$\,$hr (two-peaked lightcurve).  This period is in close
agreement with $P$ = 3.9154$\pm$0.0002$\,$hr determined independently
\citep{2006ApJ...639.1238R}.
Photometry in the other filters was scaled to the $R$-band lightcurve by
subtracting the median colors $B-R=0.972$ and $R-J=0.88$, as determined from
our data.  The $B-R$ color is again in good agreement with $B-R$ =
0.969$\pm$0.030 reported by Rabinowitz et al; these authors did not measure
$R-J$.

\FigBRJLightcurve

\FigBRColorCurve

The resulting phased $B$-, $R$- and $J$-band lightcurves of EL61 are shown in
Fig.~\ref{Fig.BRJLC}. Two main features are immediately apparent from the
lightcurves. Firstly, the two peaks of the lightcurve are unequal.  The total
range (peak-to-peak) is 0.29$\pm$0.02$\,$mag.\ but the second peak is smaller
by roughly 0.08$\,$mag.  This asymmetry in the lightcurve peaks cannot be
matched by simple equilibrium shape models of the type proposed by
\citet{2006ApJ...639.1238R}, since the latter are symmetric
\citep{1969efe..book.....C}.
Secondly, we note that, in the interval roughly from 0.7 to 1.0 in rotational
phase (Fig.~\ref{Fig.BRJLC}) the $B$ data fall systematically below the $R$
data. Although small, this effect appears in measurements from three different
nights and hence we regard it as observationally secure.  The $B-R$ color
curve, computed from the data in Fig.~\ref{Fig.BRJLC}, is shown separately in
Fig.~\ref{Fig.BRColorCurve}.  There, the $R$ magnitudes were interpolated to
the rotational phases of $B$ photometry and were subtracted from the $B$ data
points. The resulting color curve was smoothed using a running median filter to
show a reddening of up to 0.035 magnitudes. The $J$ data are of lower
signal-to-noise but, in the region near 0.75 rotational phase, $R-J$ is also
redder than near the 0.25 rotational phase peak (Fig.~\ref{Fig.BRJLC}). 

We quantitatively assess the significance of the red feature in the $B-R$ color
curve by noting that, in the interval from 0.7 to 1.0 in rotational phase, 21
of the 23 consecutive phased measurements fall above the median $B-R$ for EL61.
The probability of this result is the same as the probability of obtaining at
least 21 ``tails'' in 23 tosses of an unbiased coin. Assuming a binomial
distribution, this probability is roughly $p=3.3\times10^{-5}$, corresponding
to $\sim$4$\sigma$.  Furthermore, at least 9 measurements in that same interval
lie $>3\sigma$ above the median, corresponding to $\sqrt{9}\times
3\sigma=9\sigma$.  In this sense, the redder region in
Fig.~\ref{Fig.BRColorCurve} is unlikely to be due to chance.  This, plus the
fact that the red spot is confirmed by observations on different nights
together strongly suggest that the feature is real.

Lightcurve asymmetry of the type observed in Fig.~\ref{Fig.BRJLC} could be
caused by strength-supported topography. However, the associated color
variations (Fig.~\ref{Fig.BRColorCurve}) cannot be so explained.  Instead,
the data are best explained by the presence of wavelength-dependent albedo
markings on the surface of EL61, perhaps analogous to the ones already mapped
on Pluto.  Specifically,  given that the body shape of EL61 is close to a
figure of equilibrium, the multi-wavelength lightcurve data show the existence
of a region, near the second peak in Fig.~\ref{Fig.JacSpotCurves}, that is
darker and redder than elsewhere.  For want of a better label, we refer to this
as the ``dark red spot'' (DRS).

\FigSpotCurves

The time-resolved measurements of the 1.5$\,\mu$m water-ice band from UT 2007
July 07 are plotted in Fig.~\ref{Fig.WaterBandDepth}.  The data provide no
compelling indication of variability, except that the ratio of the 1.6$\,\mu$m
flux density to the continuum flux density at 1.25$\,\mu$m appears lower (the
water-ice band deepens) near phase $\sim$0 than at other phases.  An attempt to
repeat the 1.6$\,\mu$m photometry on UT 2007 July 08 was thwarted by unstable
atmospheric opacity.  In the absence of confirming data from a second night, we
regard the variation seen in Fig.~\ref{Fig.WaterBandDepth} as interesting but
inconclusive.  We cannot determine whether the water-band depth varies with the
rotation of EL61.

What constraints can be set on the albedo markings present on the surface of
EL61?  We first address the spatial extent of the DRS.  In principle, the DRS
could be very small compared to the instantaneous projected cross-section of
EL61 but would then need to be very red and very dark relative to the
surroundings in order to give rise to the observed lightcurve differences.  At
the other extreme, the DRS could be large, possibly even hemispheric in extent,
in which case its albedo and color contrast relative to the surroundings would
be minimal (see Fig.~\ref{Fig.SpotModels}).  To explore the range of
possibilities we computed models in which the area of the surface of EL61
occupied by the DRS was taken as a free parameter. 

The models we used are described in detail in \citet{2007AJ....133.1393L}. In
short, we render 3-dimensional models of EL61 at different rotational phases
which are used to generate the synthetic lightcurve. In this paper we adopt a
Lambert scattering law, appropriate for high-albedo icy surfaces. The spot was
simulated as a region of different reflectivity and color curves were generated
by subtracting the lightcurves of two spots of equal sizes but different
reflectivities. The shape of EL61 was modeled by a Jacobi ellipsoid with axis
ratios $b/a=0.87$ and $c/a=0.54$, which provides the best match to the
lightcurve data if no albedo variegation is present (see
Fig.~\ref{Fig.JacSpotCurves}).  The size of the spot is parametrized by its
sky-plane cross-section area relative to the maximum cross-section of the
Jacobi ellipsoid, $\pi ac$. We assumed that the rotation axis of EL61 was
inclined relative to the line-of-sight by 90$^{\circ}$, consistent with the
large measured rotational lightcurve range,  and that the DRS is located on the
equator of EL61. The observed sequence of brighter and fainter extrema
indicates that longitude of the DRS must lie in a leading quadrant with respect
to one of the semi-major axes. This prediction is corroborated by our models,
which further show that a longitudinal separation of 45\degr between the DRS
and the long axis of EL61 produces a better match to the data than 30\degr or
60\degr. The ability to fit the shape of the lightcurves in different filters
was used as a metric for the models.  Three of the best-fit examples are shown
in Figs.~\ref{Fig.JacSpotCurves} and \ref{Fig.SpotModels}. 

\FigWaterBandDepth

\FigSpotModels

The results, which confirm the qualitative expectations outlined above, are
shown in Fig.~\ref{Fig.AlbedoAndColorVsArea}. Figure~\ref{Fig.ColorVsAlbedo}
combines the color and albedo constraints and allows comparison with real
surfaces. Also marked in Fig.~\ref{Fig.ColorVsAlbedo} are the ranges of color
and albedo for established outer Solar system materials, including the dark
regions on Pluto and Saturn's satellite Iapetus.  The EL61 data are
incompatible with very small patches of dark, red matter like that found on the
low-albedo side of Iapetus, or even with the darker material on the surface of
Pluto.  Indeed, if the spot is to have a $B-R$ color within the range observed
for Solar system objects ($B-R\lesssim2$) then it must be larger than
$\sim$20\% of the maximum cross-section of EL61 (see
Fig.~\ref{Fig.AlbedoAndColorVsArea}).  Instead, Iapetus' and Pluto's bright
areas match the DRS in term of albedo and $B-R$ color. The surfaces of Eris (a
large KBO) and 2005 FY9 are also consistent with the DRS, even if these objects
have highly uncertain albedos \citep{2007astro.ph..2538S}. 
All matching surface types would imply a DRS size of 35\% to 50\%.

Another possibility is that the DRS is simply terrain contaminated by dirt.
This would account for both the darkening and the reddening, but the suspected
deepening of the 1.5$\,\mu$m water band close to the DRS in rotational phase
(see Fig.~\ref{Fig.WaterBandDepth}) would be harder to explain; a weaker, i.e.
less deep, water feature would be expected if that were the case.
Alternatively, the DRS could be a region depleted in a spectrally neutral
substance, both brighter and bluer than water ice.  A more contrived
explanation involves the presence of larger water-ice grains on the DRS which
would lower the albedo and redden the surface \citep{1982Icar...49..244C}, and
produce deeper water-ice absorption bands \citep{2005LPI....36.2061S}. On
Enceladus, larger grains are found on the region often referred to as the tiger
stripes, where cryovolcanism is thought to be happening
\citep{2007LPI....38.1747S}.

\FigAlbedoAndColorVsArea

\FigColorVsAlbedo

What might be the origin of the DRS?  On Pluto, the light and dark albedo
markings may be self-sustaining and caused by the mobility of surface
volatiles, partly driven by the seasons (Hansen and Paige 1996).  There, dark
regions are heated by the Sun leading to higher sublimation rates and the
migration of volatiles towards brighter, cooler regions.  In this way the
volatile ices may naturally migrate to restricted regions of the surface.  The
dominant volatile species on Pluto is the highly volatile solid nitrogen,
N$_2$, with methane (CH$_4$) mixed-in as an optically active tracer.  In
contrast, the surface of EL61 appears to be water-ice dominated, with no
evidence for the diagnostic $N_2$ band at 2.15 $\mu$m (Fig.
\ref{Fig.Spectrum}).  Water ice is utterly refractory at the $\sim$30 K
temperatures on EL61, and this albedo instability mechanism seems unlikely to
apply.   It has been suggested that EL61 is the source of an impact-produced
dynamical family of water-rich KBOs.  It is tempting to speculate that the DRS
could mark the scar of the impact from which the family members were
purportedly excavated, although such an explanation could hardly be unique.

\section{Summary}

From time-resolved, high precision optical and near-infared photometry of KBO
2003$\,$EL$_{61}$ we find the following results.

\begin{myitemize}

\item  \raggedright The $R$-band lightcurve has period 3.9155$\pm$0.0001$\,$hrs and
peak-to-peak range 0.29$\pm$0.02 mag.  However, successive lightcurve peaks in
the $R$-band data are clearly unequal.  The $B-R$ and $R-J$ colors of EL61 also
vary with rotational phase.  

\item No variation in the 1.5$\,\mu$m water-ice band with rotational phase
larger than $\sim$5\% is observed in our data.

\item The observed lightcurve variations are broadly consistent with a
rotational equilibrium (strengthless body) model but with the additional requirement
that the surface must support wavelength-dependent albedo variations
(``spots'') in order to explain the color variations.  

\item We explored the range of parameters of possible surface spots that are
consistent with the time-resolved photometry.  Very small ``spots'' having
albedo and color very different from the surroundings are ruled out by our
data.  Instead, the surface feature responsible for the wavelength-dependence
of the lightcurve must have an areal extent corresponding to a significant
fraction of the instantaneous projected cross-section.  

\end{myitemize}

\section*{Acknowledgments}

We thank Roy Gal and Toshi Kasuga for allowing us to observe during their
scheduled time.  Andrew Wang, Colin Aspin and John Dvorak provided invaluable
support. We appreciate comments from Jan Kleyna. PL was supported by a grant to
DJ from the National Science Foundation and NP by the Funda{\c c}\~ao para a
Ci\^encia e a Tecnologia, BPD/2004/18729 and by a grant to DJ from the NASA
Origins program.


\clearpage
\vfill

\LongTables
\begin{deluxetable}{ccc|ccc}
  \tablecaption{$B$-band Photometry. \label{Table.BData}}
   \tablewidth{0pt}
   \tablehead{
   \colhead{UT Date\tna} & \colhead{Julian Date\tna} & \colhead{$m_B$\tnb} &
   \colhead{UT Date\tna} & \colhead{Julian Date\tna} & \colhead{$m_B$\tnb}  
   }
   \startdata
     2007 Jun 14.98390 & 2454266.483896 & 18.169$\pm$0.005 & 2007 Jun 15.12433 & 2454266.624326 & 18.295$\pm$0.005\\
     2007 Jun 14.98989 & 2454266.489892 & 18.192$\pm$0.005 & 2007 Jun 15.12788 & 2454266.627878 & 18.261$\pm$0.005\\
     2007 Jun 15.00129 & 2454266.501292 & 18.287$\pm$0.005 & 2007 Jun 15.13144 & 2454266.631443 & 18.225$\pm$0.005\\
     2007 Jun 15.00572 & 2454266.505724 & 18.329$\pm$0.005 & 2007 Jun 15.13499 & 2454266.634995 & 18.202$\pm$0.005\\
     2007 Jun 15.00999 & 2454266.509995 & 18.370$\pm$0.005 & 2007 Jun 15.13855 & 2454266.638549 & 18.184$\pm$0.005\\
     2007 Jun 15.01429 & 2454266.514289 & 18.397$\pm$0.005 & 2007 Jun 15.14281 & 2454266.642808 & 18.161$\pm$0.005\\
     2007 Jun 15.01869 & 2454266.518686 & 18.403$\pm$0.005 & 2007 Jul 22.96857 & 2454304.468573 & 18.325$\pm$0.007\\
     2007 Jun 15.03386 & 2454266.533860 & 18.363$\pm$0.005 & 2007 Jul 22.97550 & 2454304.475497 & 18.283$\pm$0.007\\
     2007 Jun 15.03757 & 2454266.537575 & 18.337$\pm$0.005 & 2007 Jul 22.98246 & 2454304.482463 & 18.221$\pm$0.007\\
     2007 Jun 15.04263 & 2454266.542632 & 18.288$\pm$0.005 & 2007 Jul 22.98775 & 2454304.487747 & 18.183$\pm$0.007\\
     2007 Jun 15.04688 & 2454266.546879 & 18.235$\pm$0.005 & 2007 Jul 22.99173 & 2454304.491731 & 18.177$\pm$0.007\\
     2007 Jun 15.05056 & 2454266.550560 & 18.192$\pm$0.005 & 2007 Jul 22.99578 & 2454304.495777 & 18.177$\pm$0.007\\
     2007 Jun 15.05445 & 2454266.554448 & 18.166$\pm$0.005 & 2007 Jul 23.00453 & 2454304.504533 & 18.215$\pm$0.007\\
     2007 Jun 15.05819 & 2454266.558186 & 18.135$\pm$0.005 & 2007 Jul 23.00859 & 2454304.508593 & 18.238$\pm$0.007\\
     2007 Jun 15.06184 & 2454266.561844 & 18.114$\pm$0.005 & 2007 Jul 23.01263 & 2454304.512634 & 18.288$\pm$0.007\\
     2007 Jun 15.06554 & 2454266.565535 & 18.104$\pm$0.005 & 2007 Jul 23.01670 & 2454304.516700 & 18.328$\pm$0.007\\
     2007 Jun 15.06924 & 2454266.569239 & 18.116$\pm$0.005 & 2007 Jul 23.02072 & 2454304.520717 & 18.349$\pm$0.007\\
     2007 Jun 15.07287 & 2454266.572872 & 18.122$\pm$0.005 & 2007 Jul 23.02722 & 2454304.527220 & 18.388$\pm$0.007\\
     2007 Jun 15.07658 & 2454266.576576 & 18.148$\pm$0.005 & 2007 Jul 23.03125 & 2454304.531245 & 18.410$\pm$0.007\\
     2007 Jun 15.08022 & 2454266.580221 & 18.166$\pm$0.005 & 2007 Jul 24.97186 & 2454306.471861 & 18.283$\pm$0.014\\
     2007 Jun 15.08414 & 2454266.584144 & 18.209$\pm$0.005 & 2007 Jul 24.97540 & 2454306.475403 & 18.333$\pm$0.014\\
     2007 Jun 15.08831 & 2454266.588311 & 18.263$\pm$0.005 & 2007 Jul 24.98242 & 2454306.482416 & 18.360$\pm$0.014\\
     2007 Jun 15.09185 & 2454266.591852 & 18.283$\pm$0.005 & 2007 Jul 24.98596 & 2454306.485957 & 18.389$\pm$0.014\\
     2007 Jun 15.09541 & 2454266.595405 & 18.315$\pm$0.005 & 2007 Jul 24.99282 & 2454306.492821 & 18.400$\pm$0.014\\
     2007 Jun 15.09896 & 2454266.598958 & 18.328$\pm$0.005 & 2007 Jul 24.99636 & 2454306.496361 & 18.394$\pm$0.014\\
     2007 Jun 15.10251 & 2454266.602511 & 18.348$\pm$0.005 & 2007 Jul 25.00914 & 2454306.509139 & 18.326$\pm$0.014\\
     2007 Jun 15.10633 & 2454266.606329 & 18.356$\pm$0.005 & 2007 Jul 25.01268 & 2454306.512680 & 18.269$\pm$0.014\\
     2007 Jun 15.10989 & 2454266.609894 & 18.347$\pm$0.005 & 2007 Jul 25.01957 & 2454306.519566 & 18.233$\pm$0.014\\
     2007 Jun 15.11345 & 2454266.613448 & 18.343$\pm$0.005 & 2007 Jul 25.02311 & 2454306.523107 & 18.191$\pm$0.014\\
     2007 Jun 15.11699 & 2454266.616988 & 18.329$\pm$0.005 & 2007 Jul 25.03137 & 2454306.531371 & 18.108$\pm$0.014\\
     2007 Jun 15.12054 & 2454266.620541 & 18.320$\pm$0.005 &                   &                &                 \\
   \enddata
  \tablenotetext{a}{Dates are light-time corrected;}
  \tablenotetext{b}{Apparent magnitude.}
\end{deluxetable}

\begin{deluxetable}{ccc|ccc}
  \tablecaption{$R$-band Photometry. \label{Table.RData}}
   \tablewidth{0pt}
   \tablehead{
   \colhead{UT Date\tna} & \colhead{Julian Date\tna} & \colhead{$m_R$\tnb} & \colhead{UT Date\tna} & \colhead{Julian Date\tna} & \colhead{$m_R$\tnb}  
   }
   \startdata
     2007 Jun 10.96450 & 2454262.464497 & 17.314$\pm$0.009 &  2007 Jun 11.15618 & 2454262.656183 & 17.140$\pm$0.009\\
     2007 Jun 10.96670 & 2454262.466696 & 17.293$\pm$0.009 &  2007 Jun 11.15764 & 2454262.657642 & 17.151$\pm$0.009\\
     2007 Jun 10.96809 & 2454262.468085 & 17.280$\pm$0.009 &  2007 Jun 11.15912 & 2454262.659123 & 17.160$\pm$0.009\\
     2007 Jun 10.97004 & 2454262.470041 & 17.263$\pm$0.009 &  2007 Jun 12.96765 & 2454264.467647 & 17.291$\pm$0.006\\
     2007 Jun 10.97143 & 2454262.471430 & 17.249$\pm$0.009 &  2007 Jun 12.97079 & 2454264.470795 & 17.309$\pm$0.006\\
     2007 Jun 10.97282 & 2454262.472819 & 17.237$\pm$0.009 &  2007 Jun 12.97257 & 2454264.472566 & 17.336$\pm$0.006\\
     2007 Jun 10.97491 & 2454262.474913 & 17.212$\pm$0.009 &  2007 Jun 12.97661 & 2454264.476606 & 17.344$\pm$0.006\\
     2007 Jun 10.97637 & 2454262.476371 & 17.190$\pm$0.009 &  2007 Jun 12.97867 & 2454264.478666 & 17.362$\pm$0.006\\
     2007 Jun 10.97782 & 2454262.477818 & 17.188$\pm$0.009 &  2007 Jun 12.98030 & 2454264.480298 & 17.369$\pm$0.006\\
     2007 Jun 10.97930 & 2454262.479299 & 17.176$\pm$0.009 &  2007 Jun 12.98213 & 2454264.482126 & 17.351$\pm$0.006\\
     2007 Jun 10.98076 & 2454262.480758 & 17.160$\pm$0.009 &  2007 Jun 12.98372 & 2454264.483723 & 17.390$\pm$0.006\\
     2007 Jun 11.00040 & 2454262.500399 & 17.193$\pm$0.009 &  2007 Jun 12.98544 & 2454264.485436 & 17.390$\pm$0.006\\
     2007 Jun 11.00186 & 2454262.501856 & 17.195$\pm$0.009 &  2007 Jun 12.98704 & 2454264.487044 & 17.384$\pm$0.006\\
     2007 Jun 11.00330 & 2454262.503303 & 17.209$\pm$0.009 &  2007 Jun 12.98870 & 2454264.488700 & 17.390$\pm$0.006\\
     2007 Jun 11.00476 & 2454262.504761 & 17.223$\pm$0.009 &  2007 Jun 12.99031 & 2454264.490308 & 17.386$\pm$0.006\\
     2007 Jun 11.00622 & 2454262.506220 & 17.244$\pm$0.009 &  2007 Jun 12.99192 & 2454264.491917 & 17.390$\pm$0.006\\
     2007 Jun 11.00823 & 2454262.508233 & 17.260$\pm$0.009 &  2007 Jun 12.99350 & 2454264.493502 & 17.370$\pm$0.006\\
     2007 Jun 11.00969 & 2454262.509692 & 17.281$\pm$0.009 &  2007 Jun 12.99514 & 2454264.495145 & 17.363$\pm$0.006\\
     2007 Jun 11.01115 & 2454262.511149 & 17.281$\pm$0.009 &  2007 Jun 12.99673 & 2454264.496731 & 17.367$\pm$0.006\\
     2007 Jun 11.01260 & 2454262.512596 & 17.305$\pm$0.009 &  2007 Jun 12.99839 & 2454264.498386 & 17.356$\pm$0.006\\
     2007 Jun 11.01404 & 2454262.514043 & 17.315$\pm$0.009 &  2007 Jun 12.99994 & 2454264.499937 & 17.359$\pm$0.006\\
     2007 Jun 11.01580 & 2454262.515802 & 17.328$\pm$0.009 &  2007 Jun 13.00152 & 2454264.501523 & 17.335$\pm$0.006\\
     2007 Jun 11.01727 & 2454262.517272 & 17.328$\pm$0.009 &  2007 Jun 13.00317 & 2454264.503165 & 17.332$\pm$0.006\\
     2007 Jun 11.01873 & 2454262.518730 & 17.330$\pm$0.009 &  2007 Jun 13.00480 & 2454264.504797 & 17.325$\pm$0.006\\
     2007 Jun 11.02018 & 2454262.520176 & 17.360$\pm$0.009 &  2007 Jun 13.00657 & 2454264.506568 & 17.305$\pm$0.006\\
     2007 Jun 11.02162 & 2454262.521623 & 17.354$\pm$0.009 &  2007 Jun 13.00833 & 2454264.508327 & 17.288$\pm$0.006\\
     2007 Jun 11.02323 & 2454262.523232 & 17.365$\pm$0.009 &  2007 Jun 13.00994 & 2454264.509936 & 17.268$\pm$0.006\\
     2007 Jun 11.02468 & 2454262.524678 & 17.368$\pm$0.009 &  2007 Jun 13.01190 & 2454264.511904 & 17.251$\pm$0.006\\
     2007 Jun 11.02615 & 2454262.526148 & 17.371$\pm$0.009 &  2007 Jun 13.01366 & 2454264.513662 & 17.248$\pm$0.006\\
     2007 Jun 11.02761 & 2454262.527607 & 17.373$\pm$0.009 &  2007 Jun 13.01527 & 2454264.515271 & 17.229$\pm$0.006\\
     2007 Jun 11.02905 & 2454262.529052 & 17.378$\pm$0.009 &  2007 Jun 13.01684 & 2454264.516845 & 17.221$\pm$0.006\\
     2007 Jun 11.03132 & 2454262.531321 & 17.381$\pm$0.009 &  2007 Jun 13.01844 & 2454264.518442 & 17.202$\pm$0.006\\
     2007 Jun 11.03277 & 2454262.532768 & 17.373$\pm$0.009 &  2007 Jun 13.02004 & 2454264.520039 & 17.196$\pm$0.006\\
     2007 Jun 11.03423 & 2454262.534226 & 17.365$\pm$0.009 &  2007 Jun 13.02191 & 2454264.521914 & 17.195$\pm$0.006\\
     2007 Jun 11.03570 & 2454262.535696 & 17.381$\pm$0.009 &  2007 Jun 13.02344 & 2454264.523441 & 17.189$\pm$0.006\\
     2007 Jun 11.03714 & 2454262.537143 & 17.367$\pm$0.009 &  2007 Jun 13.02503 & 2454264.525027 & 17.181$\pm$0.006\\
     2007 Jun 11.06471 & 2454262.564711 & 17.194$\pm$0.009 &  2007 Jun 13.02662 & 2454264.526624 & 17.182$\pm$0.006\\
     2007 Jun 11.06617 & 2454262.566169 & 17.182$\pm$0.009 &  2007 Jun 13.02829 & 2454264.528290 & 17.189$\pm$0.006\\
     2007 Jun 11.06763 & 2454262.567628 & 17.184$\pm$0.009 &  2007 Jun 13.08447 & 2454264.584469 & 17.329$\pm$0.006\\
     2007 Jun 11.06913 & 2454262.569132 & 17.188$\pm$0.009 &  2007 Jun 13.08607 & 2454264.586065 & 17.301$\pm$0.006\\
     2007 Jun 11.07058 & 2454262.570579 & 17.188$\pm$0.009 &  2007 Jun 13.08778 & 2454264.587778 & 17.292$\pm$0.006\\
     2007 Jun 11.07210 & 2454262.572095 & 17.193$\pm$0.009 &  2007 Jun 13.08932 & 2454264.589318 & 17.267$\pm$0.006\\
     2007 Jun 11.07355 & 2454262.573553 & 17.197$\pm$0.009 &  2007 Jun 13.09105 & 2454264.591054 & 17.264$\pm$0.006\\
     2007 Jun 11.07500 & 2454262.574999 & 17.198$\pm$0.009 &  2007 Jun 13.09261 & 2454264.592605 & 17.230$\pm$0.006\\
     2007 Jun 11.07646 & 2454262.576458 & 17.213$\pm$0.009 &  2007 Jun 13.09419 & 2454264.594190 & 17.222$\pm$0.006\\
     2007 Jun 11.07792 & 2454262.577916 & 17.225$\pm$0.009 &  2007 Jun 13.09581 & 2454264.595810 & 17.198$\pm$0.006\\
     2007 Jun 11.08353 & 2454262.583529 & 17.270$\pm$0.009 &  2007 Jun 13.09750 & 2454264.597499 & 17.184$\pm$0.006\\
     2007 Jun 11.08499 & 2454262.584988 & 17.274$\pm$0.009 &  2007 Jun 13.09912 & 2454264.599120 & 17.171$\pm$0.006\\
     2007 Jun 11.08643 & 2454262.586434 & 17.293$\pm$0.009 &  2007 Jun 13.10073 & 2454264.600729 & 17.167$\pm$0.006\\
     2007 Jun 11.08789 & 2454262.587892 & 17.307$\pm$0.009 &  2007 Jun 13.10237 & 2454264.602372 & 17.150$\pm$0.006\\
     2007 Jun 11.08934 & 2454262.589339 & 17.309$\pm$0.009 &  2007 Jun 13.10396 & 2454264.603958 & 17.146$\pm$0.006\\
     2007 Jun 11.09200 & 2454262.592001 & 17.356$\pm$0.009 &  2007 Jun 13.10552 & 2454264.605519 & 17.131$\pm$0.006\\
     2007 Jun 11.09346 & 2454262.593459 & 17.364$\pm$0.009 &  2007 Jun 13.10723 & 2454264.607232 & 17.150$\pm$0.006\\
     2007 Jun 11.09493 & 2454262.594929 & 17.366$\pm$0.009 &  2007 Jun 13.10886 & 2454264.608864 & 17.131$\pm$0.006\\
     2007 Jun 11.09637 & 2454262.596375 & 17.408$\pm$0.009 &  2007 Jun 13.11044 & 2454264.610438 & 17.137$\pm$0.006\\
     2007 Jun 11.09783 & 2454262.597833 & 17.389$\pm$0.009 &  2007 Jun 13.11207 & 2454264.612070 & 17.137$\pm$0.006\\
     2007 Jun 11.09978 & 2454262.599777 & 17.410$\pm$0.009 &  2007 Jun 13.11370 & 2454264.613702 & 17.135$\pm$0.006\\
     2007 Jun 11.10124 & 2454262.601236 & 17.419$\pm$0.009 &  2007 Jun 13.11525 & 2454264.615252 & 17.154$\pm$0.006\\
     2007 Jun 11.10272 & 2454262.602717 & 17.412$\pm$0.009 &  2007 Jun 13.11692 & 2454264.616919 & 17.159$\pm$0.006\\
     2007 Jun 11.10419 & 2454262.604187 & 17.419$\pm$0.009 &  2007 Jun 13.11848 & 2454264.618481 & 17.177$\pm$0.006\\
     2007 Jun 11.10563 & 2454262.605633 & 17.425$\pm$0.009 &  2007 Jun 13.12009 & 2454264.620090 & 17.178$\pm$0.006\\
     2007 Jun 11.10780 & 2454262.607797 & 17.430$\pm$0.009 &  2007 Jun 13.12168 & 2454264.621676 & 17.186$\pm$0.006\\
     2007 Jun 11.10924 & 2454262.609244 & 17.422$\pm$0.009 &  2007 Jun 13.12326 & 2454264.623261 & 17.213$\pm$0.006\\
     2007 Jun 11.11070 & 2454262.610702 & 17.416$\pm$0.009 &  2007 Jun 13.12490 & 2454264.624904 & 17.230$\pm$0.006\\
     2007 Jun 11.11216 & 2454262.612161 & 17.427$\pm$0.009 &  2007 Jun 13.12654 & 2454264.626536 & 17.249$\pm$0.006\\
     2007 Jun 11.11361 & 2454262.613607 & 17.414$\pm$0.009 &  2007 Jun 13.12817 & 2454264.628168 & 17.234$\pm$0.006\\
     2007 Jun 11.11600 & 2454262.616002 & 17.413$\pm$0.009 &  2007 Jun 13.12981 & 2454264.629811 & 17.288$\pm$0.006\\
     2007 Jun 11.11746 & 2454262.617461 & 17.386$\pm$0.009 &  2007 Jun 13.13143 & 2454264.631432 & 17.300$\pm$0.006\\
     2007 Jun 11.11892 & 2454262.618919 & 17.360$\pm$0.009 &  2007 Jun 13.13302 & 2454264.633017 & 17.332$\pm$0.006\\
     2007 Jun 11.12038 & 2454262.620377 & 17.384$\pm$0.009 &  2007 Jun 13.13459 & 2454264.634590 & 17.326$\pm$0.006\\
     2007 Jun 11.12184 & 2454262.621836 & 17.367$\pm$0.009 &  2007 Jun 13.13633 & 2454264.636326 & 17.335$\pm$0.006\\
     2007 Jun 11.12435 & 2454262.624347 & 17.359$\pm$0.009 &  2007 Jun 13.13789 & 2454264.637889 & 17.350$\pm$0.006\\
     2007 Jun 11.12583 & 2454262.625828 & 17.338$\pm$0.009 &  2007 Jun 13.13942 & 2454264.639417 & 17.379$\pm$0.006\\
     2007 Jun 11.12729 & 2454262.627286 & 17.331$\pm$0.009 &  2007 Jun 13.14119 & 2454264.641187 & 17.372$\pm$0.006\\
     2007 Jun 11.12874 & 2454262.628744 & 17.316$\pm$0.009 &  2007 Jun 13.14290 & 2454264.642900 & 17.383$\pm$0.006\\
     2007 Jun 11.13019 & 2454262.630191 & 17.296$\pm$0.009 &  2007 Jun 13.14458 & 2454264.644578 & 17.400$\pm$0.006\\
     2007 Jun 11.13174 & 2454262.631742 & 17.267$\pm$0.009 &  2007 Jun 13.14623 & 2454264.646233 & 17.378$\pm$0.005\\
     2007 Jun 11.13320 & 2454262.633200 & 17.255$\pm$0.009 &  2007 Jun 13.14793 & 2454264.647934 & 17.381$\pm$0.005\\
     2007 Jun 11.13466 & 2454262.634658 & 17.239$\pm$0.009 &  2007 Jul 24.97919 & 2454306.479187 & 17.412$\pm$0.005\\
     2007 Jun 11.13612 & 2454262.636116 & 17.229$\pm$0.009 &  2007 Jul 24.98065 & 2454306.480645 & 17.417$\pm$0.005\\
     2007 Jun 11.14613 & 2454262.646127 & 17.154$\pm$0.009 &  2007 Jul 24.98979 & 2454306.489788 & 17.389$\pm$0.005\\
     2007 Jun 11.14843 & 2454262.648430 & 17.144$\pm$0.009 &  2007 Jul 24.99123 & 2454306.491234 & 17.367$\pm$0.005\\
     2007 Jun 11.14989 & 2454262.649888 & 17.141$\pm$0.009 &  2007 Jul 25.00570 & 2454306.505702 & 17.287$\pm$0.005\\
     2007 Jun 11.15137 & 2454262.651370 & 17.137$\pm$0.009 &  2007 Jul 25.00717 & 2454306.507171 & 17.277$\pm$0.005\\
     2007 Jun 11.15283 & 2454262.652828 & 17.138$\pm$0.009 &  2007 Jul 25.01650 & 2454306.516499 & 17.199$\pm$0.005\\
     2007 Jun 11.15427 & 2454262.654274 & 17.143$\pm$0.009 &  2007 Jul 25.01795 & 2454306.517946 & 17.170$\pm$0.005\\
   \enddata
  \tablenotetext{a}{Dates are light-time corrected;}
  \tablenotetext{b}{Apparent magnitude.}
\end{deluxetable}

\clearpage
\vfill

\begin{deluxetable}{ccc|ccc}
  \tablecaption{$J$-band Photometry. \label{Table.JData}}
   \tablewidth{0pt}
   \tablehead{
   \colhead{UT Date\tna} & \colhead{Julian Date\tna} & \colhead{$m_J$\tnb} & \colhead{UT Date\tna} & \colhead{Julian Date\tna} & \colhead{$m_J$\tnb}  
   }
   \startdata
     2007 Jul  6.97424 & 2454288.474241 & 16.50$\pm$0.04 & 2007 Jul  7.07745 & 2454288.577448 & 16.32$\pm$0.03\\
     2007 Jul  6.98283 & 2454288.482830 & 16.47$\pm$0.04 & 2007 Jul  7.98350 & 2454289.483504 & 16.27$\pm$0.04\\
     2007 Jul  6.99100 & 2454288.490998 & 16.38$\pm$0.03 & 2007 Jul  7.99131 & 2454289.491311 & 16.28$\pm$0.04\\
     2007 Jul  6.99918 & 2454288.499178 & 16.33$\pm$0.03 & 2007 Jul  7.99912 & 2454289.499119 & 16.34$\pm$0.04\\
     2007 Jul  7.00775 & 2454288.507750 & 16.29$\pm$0.03 & 2007 Jul  8.00692 & 2454289.506925 & 16.44$\pm$0.04\\
     2007 Jul  7.01586 & 2454288.515861 & 16.31$\pm$0.03 & 2007 Jul  8.02456 & 2454289.524560 & 16.54$\pm$0.04\\
     2007 Jul  7.02574 & 2454288.525740 & 16.42$\pm$0.03 & 2007 Jul  8.03236 & 2454289.532361 & 16.55$\pm$0.04\\
     2007 Jul  7.03704 & 2454288.537044 & 16.55$\pm$0.04 & 2007 Jul  8.04017 & 2454289.540167 & 16.47$\pm$0.04\\
     2007 Jul  7.04504 & 2454288.545038 & 16.57$\pm$0.04 & 2007 Jul  8.04800 & 2454289.547996 & 16.42$\pm$0.04\\
     2007 Jul  7.05309 & 2454288.553091 & 16.55$\pm$0.04 & 2007 Jul  8.05583 & 2454289.555835 & 16.33$\pm$0.04\\
     2007 Jul  7.06125 & 2454288.561255 & 16.49$\pm$0.03 & 2007 Jul  8.06367 & 2454289.563669 & 16.27$\pm$0.04\\
     2007 Jul  7.06936 & 2454288.569363 & 16.41$\pm$0.03 & 2007 Jul  8.07147 & 2454289.571465 & 16.27$\pm$0.04\\
   \enddata
  \tablenotetext{a}{Dates are light-time corrected;}
  \tablenotetext{b}{Apparent magnitude.}
\end{deluxetable}

\begin{deluxetable}{ccc|ccc}
  \tablecaption{Ratio of flux densities at
1.6$\,\mu$m and 1.25$\,\mu$m. \label{Table.wData}}
   \tablewidth{0pt}
   \tablehead{
   \colhead{UT Date\tna} & \colhead{Julian Date\tna} & \colhead{$f_{1.6}/f_{1.25}$\tnb} & \colhead{UT Date\tna} & \colhead{Julian Date\tna} & \colhead{$f_{1.6}/f_{1.25}$\tnb}  
   }
   \startdata
     2007 Jul  6.94315 & 2454288.443153 &  0.71$\pm$0.04  &   2007 Jul  6.94707 & 2454288.447075 &  0.71$\pm$0.04\\
     2007 Jul  6.95097 & 2454288.450974 &  0.69$\pm$0.04  &   2007 Jul  6.95489 & 2454288.454894 &  0.67$\pm$0.03\\
     2007 Jul  6.95881 & 2454288.458810 &  0.67$\pm$0.03  &   2007 Jul  6.97032 & 2454288.470325 &  0.69$\pm$0.02\\
     2007 Jul  6.97424 & 2454288.474241 &  0.68$\pm$0.02  &   2007 Jul  6.97888 & 2454288.478877 &  0.68$\pm$0.02\\
     2007 Jul  6.98283 & 2454288.482830 &  0.68$\pm$0.02  &   2007 Jul  6.98708 & 2454288.487075 &  0.67$\pm$0.02\\
     2007 Jul  6.99100 & 2454288.490998 &  0.67$\pm$0.02  &   2007 Jul  6.99527 & 2454288.495273 &  0.68$\pm$0.02\\
     2007 Jul  6.99918 & 2454288.499178 &  0.69$\pm$0.02  &   2007 Jul  7.00384 & 2454288.503841 &  0.69$\pm$0.02\\
     2007 Jul  7.00775 & 2454288.507750 &  0.69$\pm$0.02  &   2007 Jul  7.01194 & 2454288.511938 &  0.68$\pm$0.02\\
     2007 Jul  7.01586 & 2454288.515861 &  0.68$\pm$0.02  &   2007 Jul  7.02182 & 2454288.521819 &  0.67$\pm$0.02\\
     2007 Jul  7.02574 & 2454288.525740 &  0.67$\pm$0.02  &   2007 Jul  7.03314 & 2454288.533136 &  0.66$\pm$0.02\\
     2007 Jul  7.03704 & 2454288.537044 &  0.65$\pm$0.02  &   2007 Jul  7.04112 & 2454288.541120 &  0.63$\pm$0.02\\
     2007 Jul  7.04504 & 2454288.545039 &  0.64$\pm$0.02  &   2007 Jul  7.04918 & 2454288.549176 &  0.66$\pm$0.02\\
     2007 Jul  7.05309 & 2454288.553091 &  0.64$\pm$0.02  &   2007 Jul  7.05733 & 2454288.557332 &  0.63$\pm$0.02\\
     2007 Jul  7.06125 & 2454288.561255 &  0.62$\pm$0.02  &   2007 Jul  7.06544 & 2454288.565437 &  0.63$\pm$0.02\\
     2007 Jul  7.06936 & 2454288.569363 &  0.64$\pm$0.02  &   2007 Jul  7.07356 & 2454288.573560 &  0.65$\pm$0.02\\
     2007 Jul  7.07745 & 2454288.577448 &  0.66$\pm$0.02  &   2007 Jul  7.08179 & 2454288.581790 &  0.68$\pm$0.02\\
     2007 Jul  6.93925 & 2454288.439254 &  0.69$\pm$0.04  &                     &                &               \\
   \enddata
  \tablenotetext{a}{Dates are light-time corrected;}
  \tablenotetext{b}{Ratio of the flux density at 1.6$\,\mu$m to the flux density at 1.25$\,\mu$m.}
\end{deluxetable}

\end{document}